# A perspective on anyonic braiding statistics

*Nicholas Read* [1,2] *and Sankar Das Sarma* [3]

[1] Department of Physics, Yale University, P.O. Box 208120, New Haven, CT 06520, USA
[2] Department of Applied Physics, Yale University, P.O. Box 208284, New Haven, CT 06520, USA
[3] Condensed Matter Theory Center, Department of Physics, University of Maryland, College Park, MD 20742, USA

In a recent paper [1], the authors performed a ground-breaking experiment, using Fabry-Perot interferometry, involving the fractionally-charged quasiparticles in the fractional quantum Hall effect (FQHE) at filling factor 1/3. In particular, the results provide strong evidence that the quasiparticles obey fractional exchange (or braiding) statistics (i.e. are anyons), as was long predicted [2]. Further, the measured quantity $\theta_a$ (= $\theta_{anyon}$, and is defined below) agrees with the theoretical prediction, $\theta_a = 2\pi/3$ (modulo $2\pi$), within experimental accuracy. As we will explain, there is a related theoretical quantity, $\theta$, such that $\theta_a = 2\theta$ (modulo $2\pi$), so the experiment finds that, modulo $2\pi$, $\theta$ must be close either to $\pi/3$ or to $4\pi/3$. It is of theoretical interest to remove this remaining ambiguity by $\pi$, because $\theta$ fully characterizes the braiding statistics.

Anyons [3,4] have the defining property that when two identical anyons (in two spatial dimensions) are exchanged adiabatically using a non-self-intersecting counterclockwise closed path that does not enclose any other anyons (an "elementary exchange"), the quantum-mechanical wavefunction acquires a phase factor $e^{i\theta}$; $\theta$, defined modulo $2\pi$, is called the statistical phase, and characterizes the type of anyon. More generally, for *n* identical quasiparticles, the elementary exchanges and their inverses generate a group of "braids", the Artin braid group $B_n$ [5,6], and the value of $\theta$ (modulo $2\pi$) uniquely determines an irreducible representation of $B_n$, which associates to each braid a number of the form $e^{im\theta}$ (*m* an integer) [6]; this constitutes the braiding statistics. As examples, $e^{i\theta} = +1$ corresponds to bosons (where, say, $\theta = 0$, or $2\pi$, etc) while $e^{i\theta} = -1$ (where $\theta = \pi$, modulo $2\pi$) describes fermions; all other (real) values of $\theta$ (modulo $2\pi$) are considered fractional, and describe anyons [3,4].

The experiment involves interference between the amplitudes for two alternative paths along which a quasiparticle can travel; the two paths together surround a region of 1/3 state. When an additional quasiparticle is added to the region, the interference phase factor changes by $e^{i\theta_a}$, where $\theta_a = 2\theta$ (modulo $2\pi$), because the composite path encloses the added quasiparticle, and that is equivalent to repeating an elementary exchange, giving $e^{i\theta_a} = e^{2i\theta}$. The values of $e^{i\theta_a}$ and of $\theta_a$ (modulo $2\pi$) were inferred from the experiment [1]. Note that any value $e^{i\theta_a} \neq 1$ means that the quasiparticles are anyons.

In the Laughlin FQHE states at filling factors $1/q$ (*q* = 1, 3, 5, . . .) [7], the theoretical value is $e^{i\theta} = e^{i\pi/q}$, or $\theta = \pi/q$ (modulo $2\pi$) [8]. For *q* = 3, within experimental accuracy, the value $\theta_a = 2\pi/3$ (modulo $2\pi$) was obtained [1], which shows that the quasiparticles are anyons, and agrees with the quantitative prediction. However, dividing by 2 gives $\theta = \pi/3$ modulo $\pi$, not modulo $2\pi$, so $\theta$ could be either $\pi/3$ or $4\pi/3$, and is not uniquely determined modulo $2\pi$ in this particular experiment, although the anyonic nature of the quasiparticles is established. Just to emphasize, an analogous experiment could not distinguish bosons from fermions; both would give $\theta_a = 0$ (modulo $2\pi$), and indeed the authors did such a measurement for fermions ([1], Extended Data

Figure 6). (The same issue of the statistical phase being determined only modulo π, not modulo 2π, also occurs in subsequent experiments by the authors at other filling factors [9].)

This point, that the anyon statistics or statistical phase is not fully determined by a measurement of $\theta_a$, has been somewhat ignored in numerous discussions and commentaries on the experimental result, including in the News and Views piece accompanying the publication [10]. We want to emphasize that, not only is a repeated elementary exchange a braid, but an elementary exchange is also a braid, according to the definition of braids (Artin's work dates originally from the 1920s [5]), and that the fundamental theoretical quantity of interest is $\theta$ (modulo 2π), the statistical phase for an elementary exchange, or equivalently $e^{i\theta}$ [3,4,6]; this determines the braiding statistics, while $e^{2i\theta}$ determines $e^{i\theta}$ only up to a factor $\pm 1$.

In conclusion, there is no suggestion that the experiment [1] did not correctly measure $e^{i\theta_a} = e^{2i\theta}$, demonstrating that the quasiparticles are anyons, but we do want to point out that it would be of great interest to design and perform an experiment to determine $e^{i\theta}$ uniquely, not only $e^{2i\theta}$.

**Acknowledgement:** NR thanks the organizers of the Nobel Symposium on anyons held in June, 2023, at which Mike Manfra gave a very clear presentation of his group's work, and also thanks Mike for discussions that helped improve the presentation of this note.